\documentclass{iaus}
\usepackage{graphicx}

\title[Role of massive stars] 
{The role of massive stars in galactic chemical evolution}

\author[Francesca Matteucci]   
{Francesca Matteucci$^1$}

\affiliation{$^1$Dipartimento di Astronomia, Universita di Trieste \\ 
Via G.B. Tiepolo, 11\\
I-34124 Trieste, Italy \\ email: {\tt matteucci@oats.inaf.it}}

\pubyear{2008}
\volume{250}  
\pagerange{119--126}
\jname{Massive Stars as Cosmic Engines}
\editors{F. Bresolin, P.A. Crowther \& J. Puls}
\begin{document}

\maketitle

\begin{abstract}
I will review the role of massive stars in galactic evolution both from the nucleosynthesis and energetics point of view. In particular, I will highlight some important observational facts explained by means of massive stars in galaxies of different morphological type: the Milky Way, ellipticals and dwarf spheroidals. I will describe first the time-delay model and its interpretation in terms of abundance ratios in galaxies, then I will discuss the importance of mass loss in massive stars to reproduce the data in the Galactic bulge and disk.
I will discuss also how massive stars can be important producers of primary nitrogen if rotation in stellar models is taken into account. Concerning elliptical galaxies, I will show that to reproduce the observed [Mg/Fe] versus Mass relation in these galaxies it is necessary to assume a more important role of massive stars in more massive galaxies and that this can be achieved by means of downsizing in star formation. I will discuss how massive stars are responsible in triggering galactic winds both in ellipticals and dwarf sheroidals. These latter systems show a low overabundance of $\alpha$-elements relative to Fe with respect to Galactic stars of the same [Fe/H]: this is interpreted as due to a slow star formation coupled with very efficient galactic winds. Finally, I will show a comparison between the predicted Type Ib/c rates in galaxies and the observed GRB rate and how we can impose constraints on the mechanism of galaxy formation by studying the GRB rate at high redshift.

\end{abstract}

\firstsection 
\section{Introduction}

We call massive stars all stars with Main Sequence masses $M > 8 M_{\odot}$, namely those stars which do not develop a degenerate carbon-oxygen core. If their mass is lower than 10 $M_{\odot}$, they will explode as e-capture supernovae (SNe): the explosion is triggered by e-capture which destabilizes the star and then by the ignition of oxygen in a degenerate O-Ne-Mg core. All the stars with mass between 8 and 10$M_{\odot}$ will instead ignite all the nuclear fuels up to Si-burning which produces $^{56}Ni$ which then decays into $^{56}Fe$. They die as core-collapse SNe which include both Type II and Ib/c SNe. The massive stars are responsible for the production of the bulk of $\alpha$-elements (O, Mg, Ne, Si, S, Ca and Ti) plus some Fe, originating either from the inner core or during the explosion by 
means of explosive Si-burning. However, the bulk of Fe should be produced by Type Ia SNe which are believed to originate from C-O white dwarfs (WDs) in binary systems, which explode by C-deflagration when the WD reaches the Chandrasekhar mass ($\sim 1.4 M_{\odot}$). Two main paths have been identified for Type Ia SNe: i) the single-degenerate scenario, made of a WD plus a normal star, where the WD explodes after reaching the limiting mass as a consequence of accreting mass from the companion, ii) the double-degenerate scenario, where two WDs of roughly 0.7$M_{\odot}$ merge after loosing angular momentum caused by gravitational wave emission. When they merge, the Chandrasekhar limiting mass is reached and the C-deflagration, producing mainly Fe, occurs as in the single-degenerate case. 

Yields from massive stars have been calculated by many authors, here I will recall some of the most recent calculations: Nomoto et al. (2006) provided detailed yields of many isotopes including explosive nucleosynthesis for massive stars without mass loss as functions of the stellar metallicity.
On the other hand,  Hirschi (2007) presented yields of a limited number of elements without explosive nucleosynthesis but taking into account mass loss and rotation. In particular, mass loss in massive stars mainly affects the yields of He, C and O as already pointed out by Maeder (1992): stellar models with mass loss predict a larger He and C production at expenses of oxygen production in massive stars. Another interesting aspect is that stellar rotation, particularly important at low metallicities, can produce a considerable amount of primary $^{14}N$ from massive stars. We define ``primary'' a chemical element which is produced directly from H and He inside the stars, whereas we define ``secondary'' any chemical species which is produced by means of heavy elements already present in the star at birth. A typical example of secondary element is represented by $^{14}N$ which originates during the CNO cycle from the original C and O present in the star. However,  $^{14}N$ can be primary if the C and O used to form it are produced by the star in situ and this is the case during the third dredge-up acting in conjunction with hot-bottom burning in
 Asymptotic Giant Branch (AGB) stars (e.g. Renzini \& Voli 1981). Rotation in massive stars can also produce primary  $^{14}N$, as shown by Meynet \& Maeder (2002).

When stellar yields are included in a chemical evolution model, namely a model which is aimed at predicting the temporal and spatial evolution of the abundances of the most abundant isotopes in the interstellar gas, we can compare the model results with detailed and precise abundance determinations. From this comparison we can then infer important constraints both on stellar nucleosynthesis, initial mass function (IMF), history of star formation (SF) and mechanisms of galaxy formation. In this paper, we will show how we can succesfully interpret abundances in galaxies of various morphological type (Milky Way, ellipticals, dwarf spheroidals)
 by means of detailed chemical evolution models. In particular, we will highlight the role of massive stars in galactic chemical evolution both from the nucleosynthesis and energetics point of view. In fact, massive stars, besides producing the majority of heavy elements,  inject large quantities of energy into the interstellar medium (ISM) by means of SN explosions but also by means of stellar winds. Such a feedback is an extremely important ingredient in studying galaxy formation and evolution since it can trigger galactic winds which in turn eject 
the heavy elements into the intergalactic and intracluster medium.
Finally, since some Gamma Ray Bursts (GRBs) have been associated with Type Ib/c SNe, we will show a comparison between the rates of Type Ib/c SNe in galaxies and the observed GRB rate. Type Ib SNe are probably the result of the explosion of single  Wolf-Rayet stars ($M>25 M_{\odot}$), whereas Type Ic SNe should arise from the explosion of massive stars (12-20$M_{\odot}$) in binary systems (e.g. Baron 1992).  In particular, we will discuss predictions relative to the cosmic Type Ib/c SN rate in different scenarios of galaxy formation: monolithic and hierarchical, and show how the rate of GRBs at high redshift can be used to impose constraints on galaxy formation models.

\begin{figure}[b]
\begin{center}
 \includegraphics[width=3.4in]{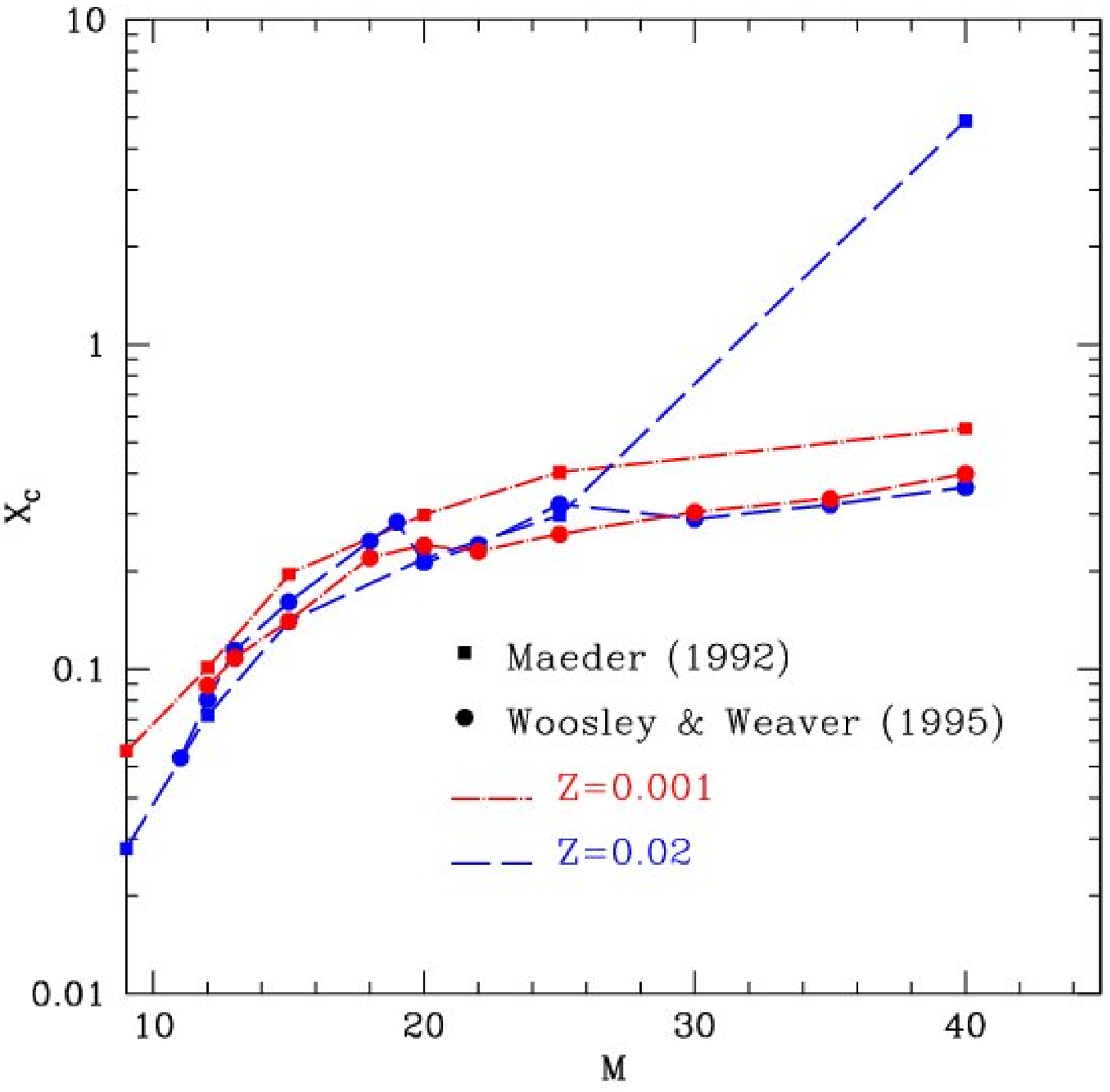} 
 \includegraphics[width=3.4in]{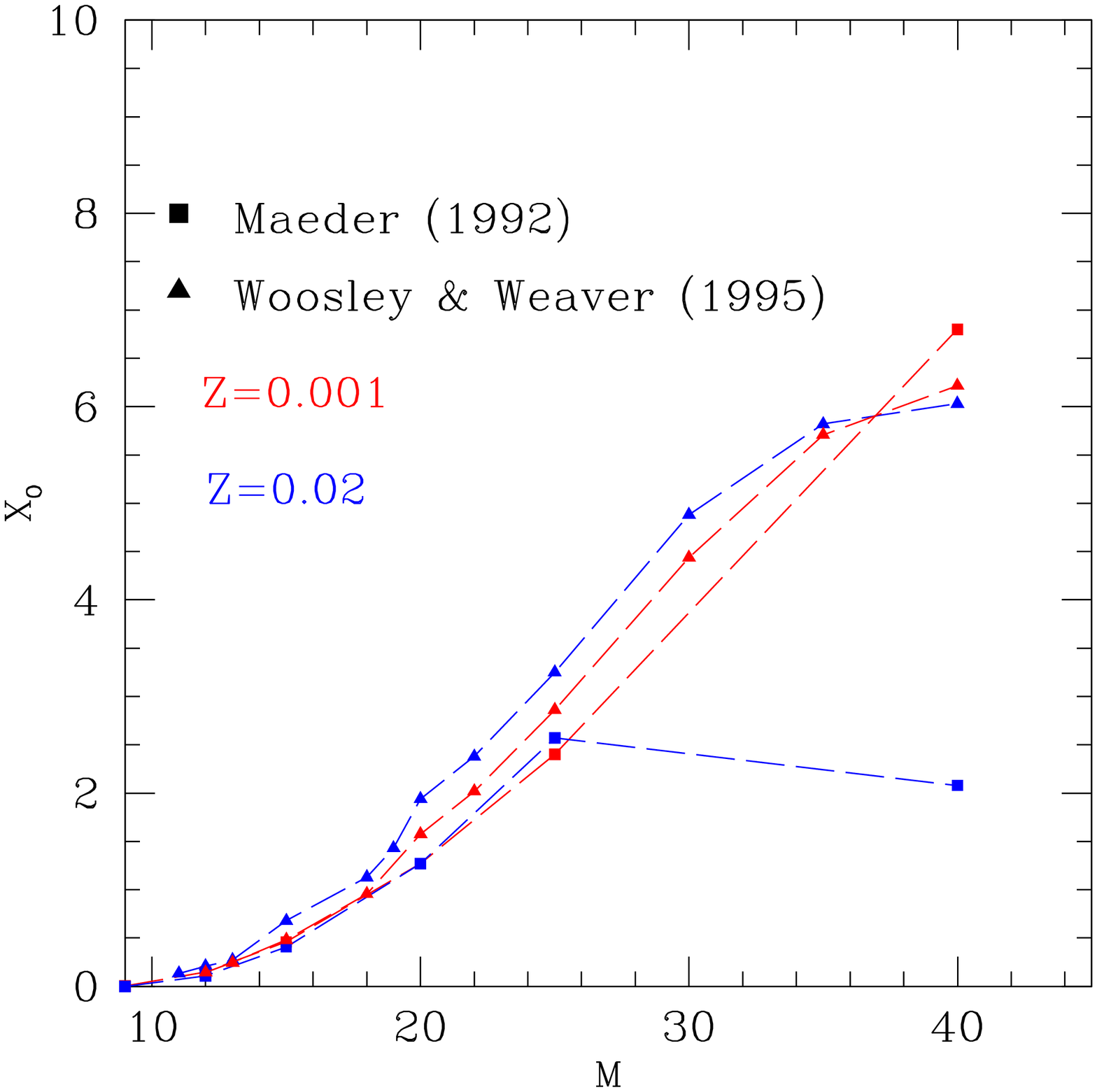} 
 \caption{Upper panel: the effect of mass loss on stellar yields: the yields of carbon from massive stars in presence of mass loss  and as functions of metallicity from Maeder (1992). The yields of carbon from the conservative models of Woosley \& Weaver (1995) are shown for comparison. Lower panel: the effect of mass loss on stellar yields. The yields of oxygen from massive stars in presence of mass loss  and as functions of metallicity from Maeder (1992). The yields of oxygen from the conservative models of Woosley \& Weaver(1995) are shown for comparison.}
   \label{fig1}
\end{center}
\end{figure}

\section{Yields from massive stars}
In Fig 1.upper panel  we show the effect of mass loss in massive stars on the yield of carbon by comparing the yields of Maeder (1992) with a high rate of mass loss with those computed by Woosley \& Weaver (1995) without mass loss. Both sets of yields are computed for two different initial stellar metallicities. As one can see, the effect of mass loss (increase in the C production) becomes evident only for a metallicity $Z\ge Z_{\odot}$ and for stars with mass $M>25 M_{\odot}$. This mass limit separates stars which end their lives as Type II SNe from those which become Wolf-Rayet stars and die as Type Ib SNe. For metallicities below the solar one the yields with and without mass loss are very similar. In Fig. 1 lower panel we show the effect of mass loss on the O yields: in this case, the O yields is severely decreased for stars with $M> 25 M_{\odot}$ and metallicities $Z \ge Z_{\odot}$. In fact, mass loss subtracts carbon to further processing through the $^{12}C(\alpha, \gamma)^{16}O$ reaction and increases with metallicity, thus its effect is evident mainly at high metallicities. 

\section{Galactic chemical evolution} 

The main ingredients to build models of galactic chemical evolution are:
\begin {itemize}

\item Initial conditions: open or closed model, primordial or pre-enriched gas.

\item  The birthrate function, in other words the star formation rate (SFR) and the initial mass function (IMF).

\item Stellar yields: newly processed and unprocessed material restored into the ISM at the star death.

\item Infall, outflow and inflow of gas.

\item Equations including all of that (e.g. Tinsley 1980, Matteucci 2001).

\end{itemize}

\subsection{The Milky Way} 
To describe our Galaxy we will assume the two-infall model proposed by Chiappini et al. (1997). In this model the Milky Way forms mainly during two main gas accretion episodes: during the first episode the halo, the central bulge and part of the thick disk form on a timescale not longer than 1-2 Gyr, whereas during a much longer second infall episode the thin disk formed inside-out. The timescale suggested for the formation of the thin disk at the solar neighbourhood is 6-8 Gyr (Chiappini et al. 1997; Boissier \& Prantzos 1999). This model assumes the Scalo (1986) IMF and a Schmidt (1959) law for the SFR with exponent $k=1.5$ and also the existence of a threshold density for the SF of $7 M_{\odot} pc^{-2}$ in the thin disk. This model reproduces the [X/Fe] vs. [Fe/H] relations found for halo and disk stars, the present time gas surface density, SFR, infall rate and SN rates. It can explain also the abundance gradients along the thin disk 
(Chiappini et al. 2001; Cescutti et al. 2007).

It is important to recall that the [X/Fe] vs. [Fe/H] relations depend mainly on the assumed nucleosynthesis, IMF and SFR. From the point of view of nucleosynthesi,s it is particularly important the role played by different SNe in the chemical enrichment. In particular, the delay with which Type Ia SNe restore the bulk of Fe relative to the fast production of $\alpha$-elements by the core-collapse SNe. This interpretation first proposed by Tinsley (1979) and then developed by Greggio \& Renzini (1983) and Matteucci \& Greggio (1986) is known as {\it time-delay model}. To illustrate the time-delay model we show in Fig. 2 upper panel 
the predicted [$\alpha$/Fe] vs. [Fe/H]  for different histories of SF.
If one assumes that O is mainly produced by Type II SNe  and that 2/3 of the total Fe is produced by Type Ia SNe whereas the remaining 1/3 is formed in massive stars, one obtains a very good fit of the data relative to the stars of the halo and disk in the solar vicinity (central curve in the figure) as well as for the bulge (upper curve) and irregular galaxies (lowest curve). On the other hand, if one assumes that Fe is either produced entirely by Type Ia SNe or
entirely by Type II SNe, the agreement with the data is lost. This simply means that both SN Types should contribute to the Fe production and that Type Ia SNe restore Fe into the ISM with a delay relative to the Fe produced by Type II SNe.
In particular, the bulk of Fe production in the solar neighbourhood occurred with a delay of $\sim$ 1 Gyr. This does not mean that the first Type Ia SNe occurred after 1Gyr, since the most massive binary systems giving rise to Type Ia SNe live no longer than 30-40 Myr. These prompt Type Ia SNe do exist,  
as shown by Mannucci et al. (2005;2006). 
\begin{figure}[b]
\begin{center}
\includegraphics[width=3.4in]{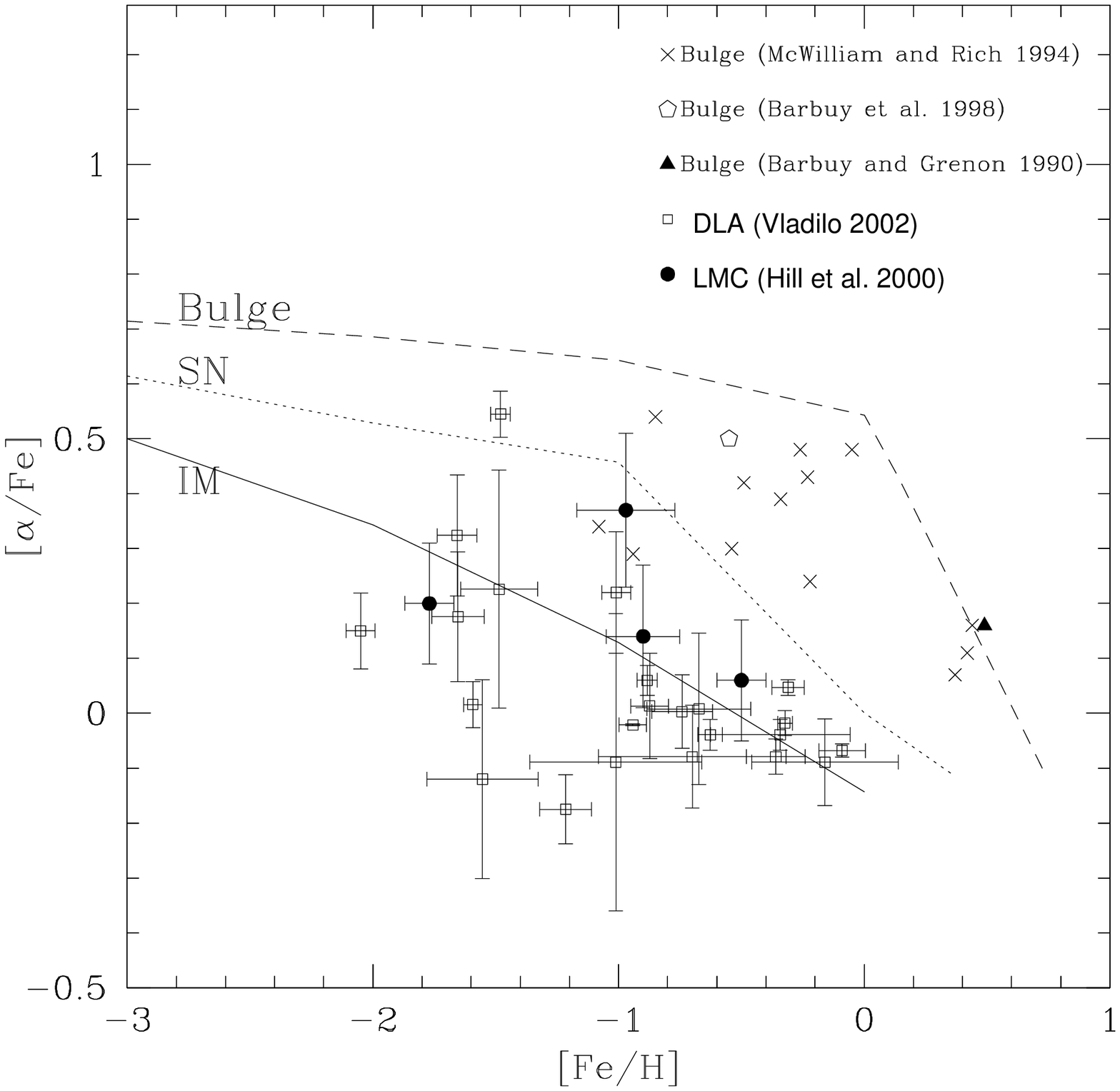} 
\includegraphics[width=4.0in]{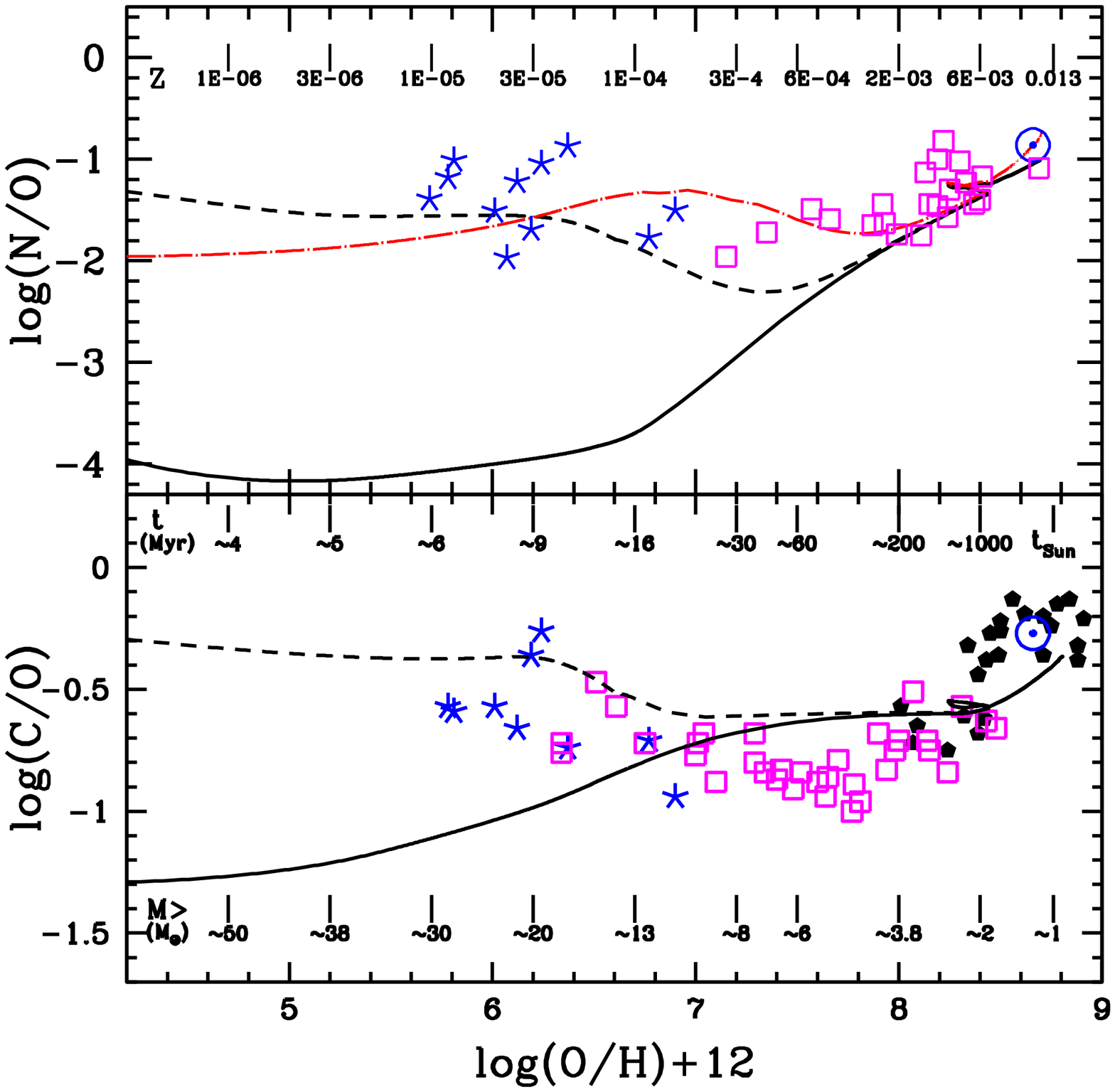} 
 \caption{Upper panel: illustration of the time-delay model for different histories of star formation.Lower panel: C and N evolution in Galactic stars. Figure and references to the data in Chiappini et al. (2006). The dashed lines represent the predictions of a model including yields from massive stars with a strong primary N production due to stellar rotation in extremely metal poor stars.}
   \label{fig1}
\end{center}
\end{figure}
However, the [$\alpha$/Fe] vs. [Fe/H] relation is not the same everywhere: in fact, even if the roles of the two different Types of SNe in producing O and Fe 
are likely to be the same and the IMF is not too different in different galaxies, the SFR is instead very different in different galaxies. In particular, the SFR must have been much faster in spheroids and ellipticals which have processed their gas in stars very quickly and at high redshift, as opposed to the slow and gasping SFR occurring in dwarf irregulars. Spirals like the Milky Way must have had a SFR intermediate and continuous. The different SFRs in galaxies influence the the age-metallicity relation, producing a very fast increase of the [Fe/H] in spheroids due mostly to Type II SNe. The opposite occurs if the SFR is slow. 
This effect is clearly shown in Fig.2 (upper panel, lowest curve).

Another important aspect of the Galactic abundance patterns is related to the N production, as shown in Fig. 2 lower panel. The most recent data seem to indicate that there should be a non negligible primary N production from massive stars. Models with stellar rotation produce N yields which can explain the observations in Galactic stars.

{\underline{\it The Galactic Bulge}} is a spheroid and its properties are more similar to those of a small elliptical than to those of the Galactic disk. Successful models for the Galactic bulge suggest that it formed very quickly during the collapse of the inner halo and that its SFR was very high like in a star-burst (Matteucci \& Brocato, 1990; Ballero et al. 2007).
In particular, the model of Ballero et al. (2007) predicts that the bulge formed on a timescale not longer than 0.1-0.5 Gyr, that the SFR was roughly 20 times more efficient than in the Galactic disk and that the IMF was flatter than that in the disk in order to reproduce the bulge stellar metallicity distribution. In Fig.3 (upper panel) we show the predicted [O/Fe] vs. [Fe/H] in the bulge compared with the most recent and accurate data. As explained before, the strong SFR acts in a way such that a solar [Fe/H] is reached in the gas before the bulk of Fe is restored from Type Ia SNe. This produces a long plateau in the [$\alpha$/Fe] ratio extending from low metallicities up to oversolar metallicities. The good agreeement between predictions and observations strongly supports the time-delay model.
\begin{figure}[b]
\begin{center}
 \includegraphics[width=3.4in]{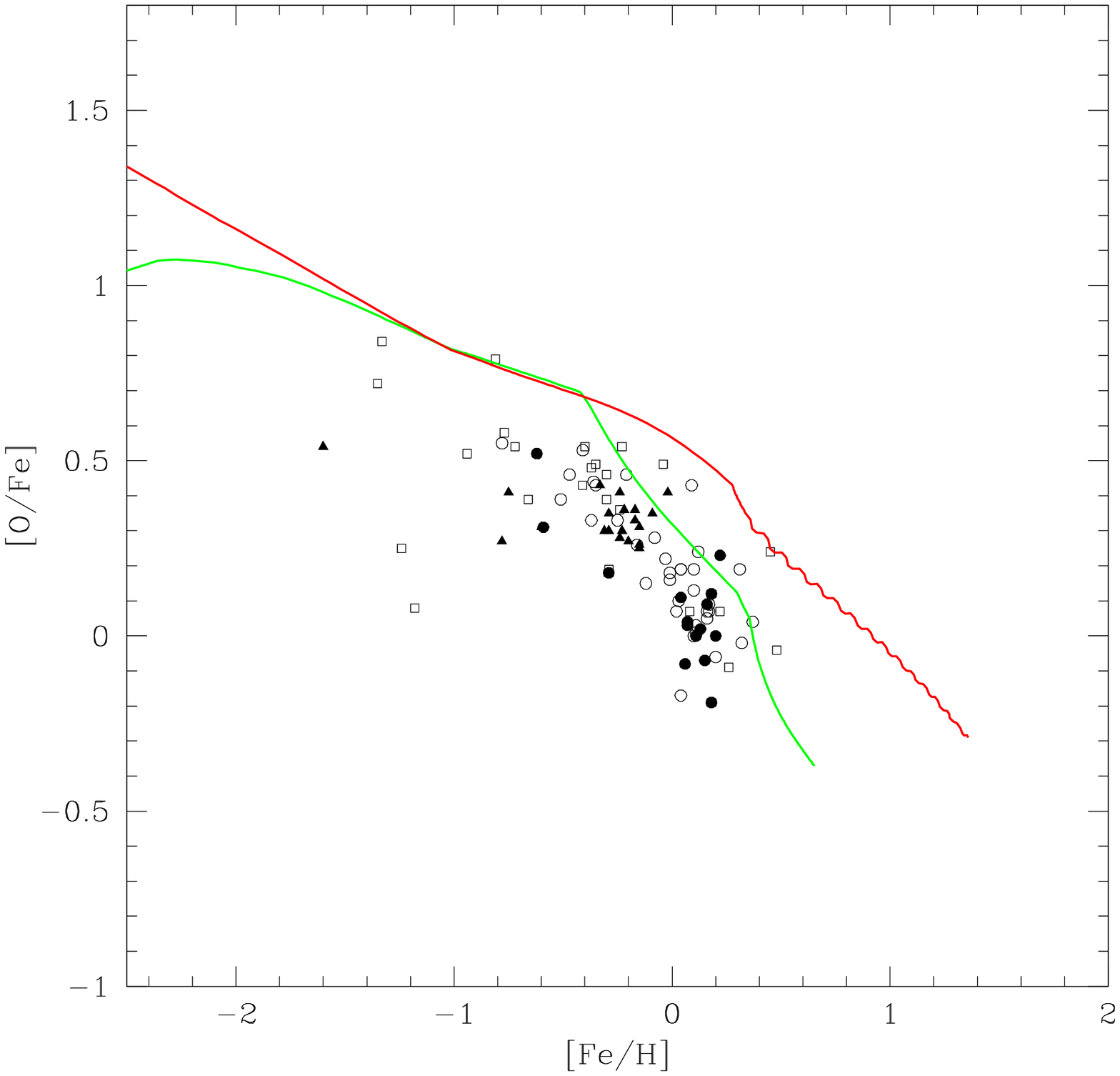} 
\includegraphics[width=3.4in]{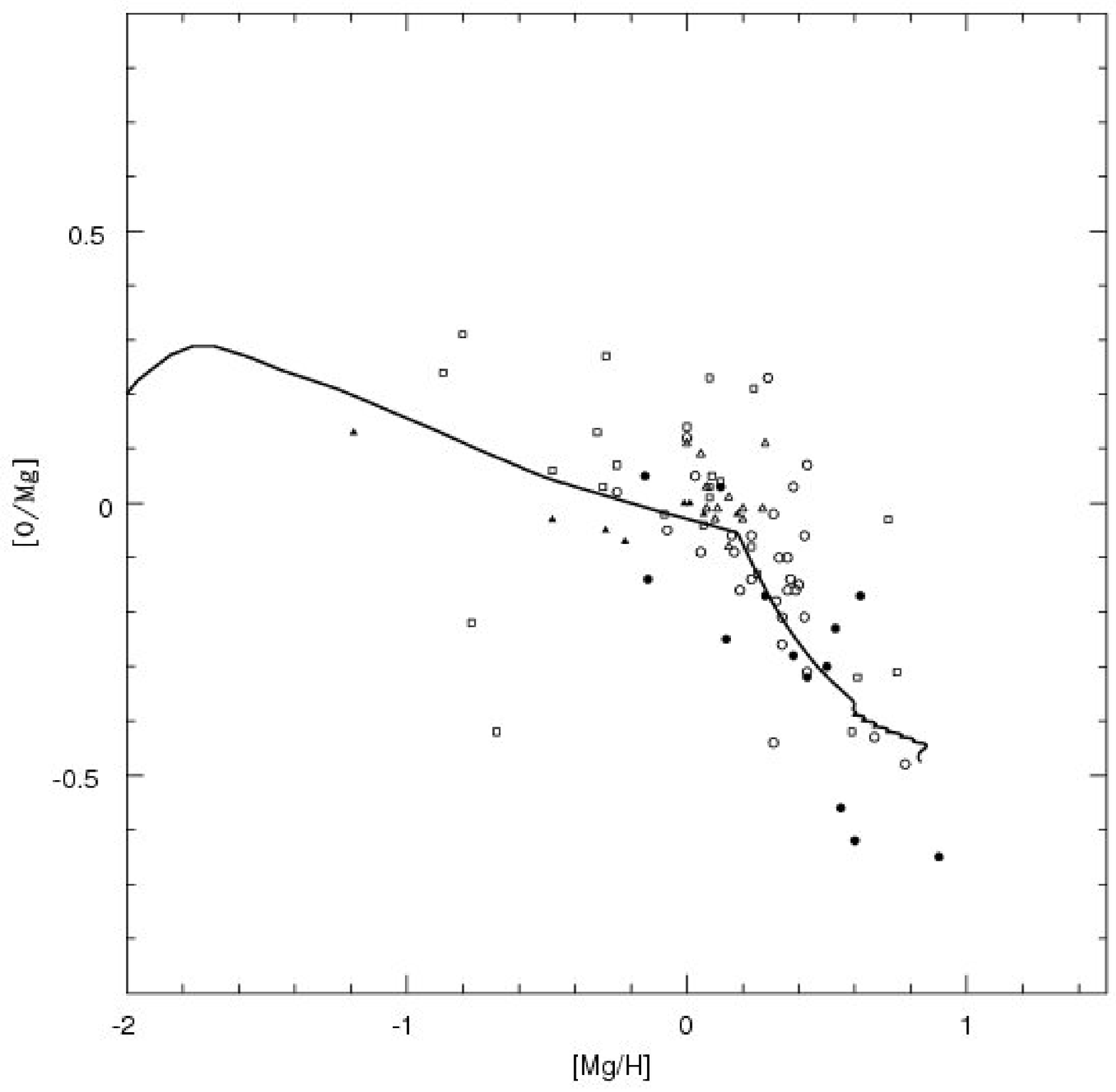} 
 \caption{Upper panel: observed and predicted [O/Fe] vs. [Fe/H]. The upper curve refers to the model with the yields of Woosley \& Weaver (1995) in massive stars, whereas the lower one to the yields of Maeder (1992). For the references on the data see McWilliam et al. (2007). Lower panel: predicted and observed [O/Mg] vs. [Mg/H] in the Galactic bulge. The model is that with the yields of Maeder (1992) for massive stars. Models adopting yields with no mass loss predict a flat [O/Mg] ratio (see McWilliam et al. (2007). }
   \label{fig1}
\end{center}
\end{figure}
In Fig. 3 upper panel are shown two model predictions: one refers to the yields of Woosley \& Weaver (1995) for massive stars whereas the other refers to a model adopting the yields of O from Maeder (1992) where mass loss depending on metallicity is taken into account. As one can see, the agreement is much better in this second case. However, as McWilliam et al. (2007) have pointed out, there is a diagram which illustrates even better the necessity of mass loss in massive stars. This is the plot in Fig. 3 lower panel, where we show the [O/Mg] vs. [Mg/H]. Here the time-delay model does not work, since both Mg and O are produced mainly in massive stars. 
In spite of this, the [O/Mg] ratio declines strongly for [Mg/H] $> 0$, a rather 
unexpected result which can be explained only if the yields of O have a strong dependence on metallicity, whereas those of Mg do not, as it is in the case of massive stars with mass loss. It is important to note that the same behaviour of the [O/Mg] ratio is shown by galactic stars (Bensby et al. 2005 ; Mc William et al. 2007).

\subsection{Elliptical galaxies} 
Early type galaxies and spheroids in general, where no gas is present now, are likely to have suffered galactic winds and/or stripping phenomena in high density environments. Galactic winds should be triggered by the energy injected by SNe and stellar winds from massive stars into the ISM. Unfortunately, there not exists a precise recipe for the feed-back and in modelling galaxy evolution one is forced to assume that a certain fraction of the  initial blast wave energy is transferred into the ISM as thermal energy. Pipino \& Matteucci (2004) have shown that if a 20 \% of the total blast wave energy is transferred from SNe into the ISM, galactic winds can occur even in massive ellipticals. However, their model does not take into account the gas cooling which can change drastically the situation. When cooling is assumed, then it is difficult to obtain galactic winds without assuming a contribution from the central AGN (e.g. Granato et al. 2001). In any case the situation is still unclear except for the fact that we know that to reproduce the observed features of local ellipticals one has to assume that their stars formed quickly at high redshift and that some mechanism must have kept the galaxies free of gas for several Gyrs. In the model of Pipino \& Matteucci (2004) the galactic winds occur quite early in the life of ellipticals and earlier in the most massive ones. In other words, in their model the most massive ellipticals form stars more intensively and for a shorter time than the 
less massive ones. This is the ``inverse wind scenario'' proposed by Matteucci (1994) which produces a downsizing in the SF.
This downsizing is responsible for the growth of the [Mg/Fe] ratio with total galactic mass observed in ellipticals and impossible to reproduce in the framework of classic hierarchical models for galaxy formation, as illustrated in Fig.4.
The effect of massive stars on the chemical evolution of ellipticals is therefore evident: it is more important in more massive objects where SF lasts for a shorter period thus creating a higher [Mg/Fe] than in less massive objects, as a consequence of the time-delay model. 
\begin{figure}[b]
\begin{center}
 \includegraphics[width=4.0in]{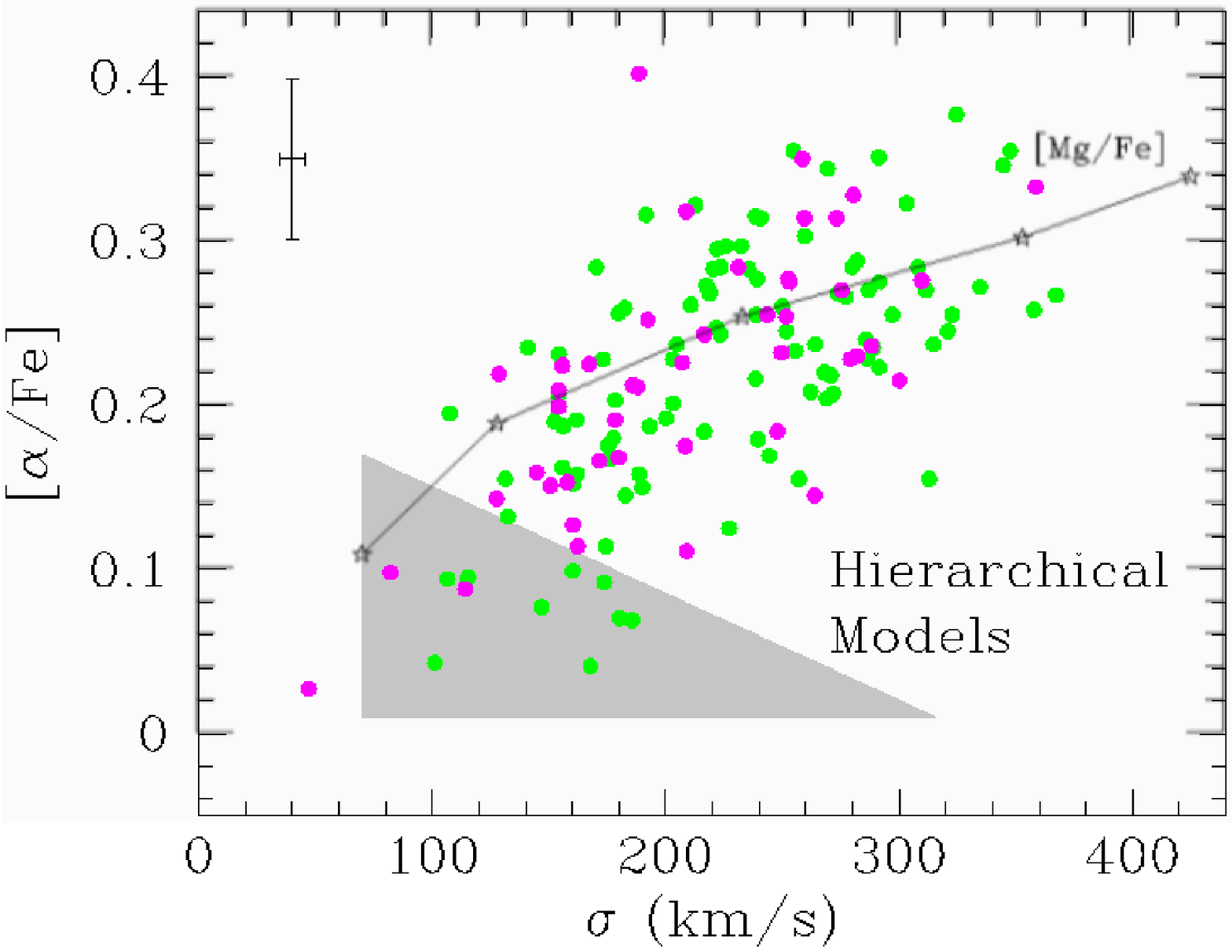} 
 \caption{Predicted and observed [$\alpha$/Fe] ratios in ellipticals. The continuous line represents the prediction of the model by Pipino \& Matteucci (2004). The shaded area represents the prediction of hierarchical models for the formation of ellipticals.The symbols are observational data. Figure adapted from Thomas et al. (2002).}
   \label{fig1}
\end{center}
\end{figure}

\subsection{Dwarf Spheroidals} 

A great deal of observational work concerning dwarf spheroidals of the Local Group has appeared in the last few years. High resolution abundance determinations allow us to compare the chemical evolution of these objects with that of the Milky Way. In Fig. 5 we show a comparison between [$\alpha$/Fe] vs. [Fe/H] relations in the dwarf spheroidals and in the Galaxy. As one can see, the observed patterns are different in the sense that except for a small overlapping of the [$\alpha$/Fe] ratios at low metallicities, generally dwarf spheroidals show lower [$\alpha$/Fe] ratios at the same [Fe/H] relative to Galactic stars. The behavior of the [$\alpha$/Fe] ratios in these objects resembles that shown in Fig. 2 upper panel for a galaxy with low star formation efficiency. By adopting a low star formation efficiency relative to the Galaxy and strong galactic winds, Lanfranchi \& Matteucci (2003, 2004) and Lanfranchi et al. (2006) reproduced the abundance patterns of six dwarf spheroidals of the Local Group both for $\alpha$-,  $s$- and $r$-process elements. They also reproduced very well the observed stellar metallicity distribution of the Carina galaxy.
\begin{figure}[b]
\begin{center}
 \includegraphics[width=3.4in]{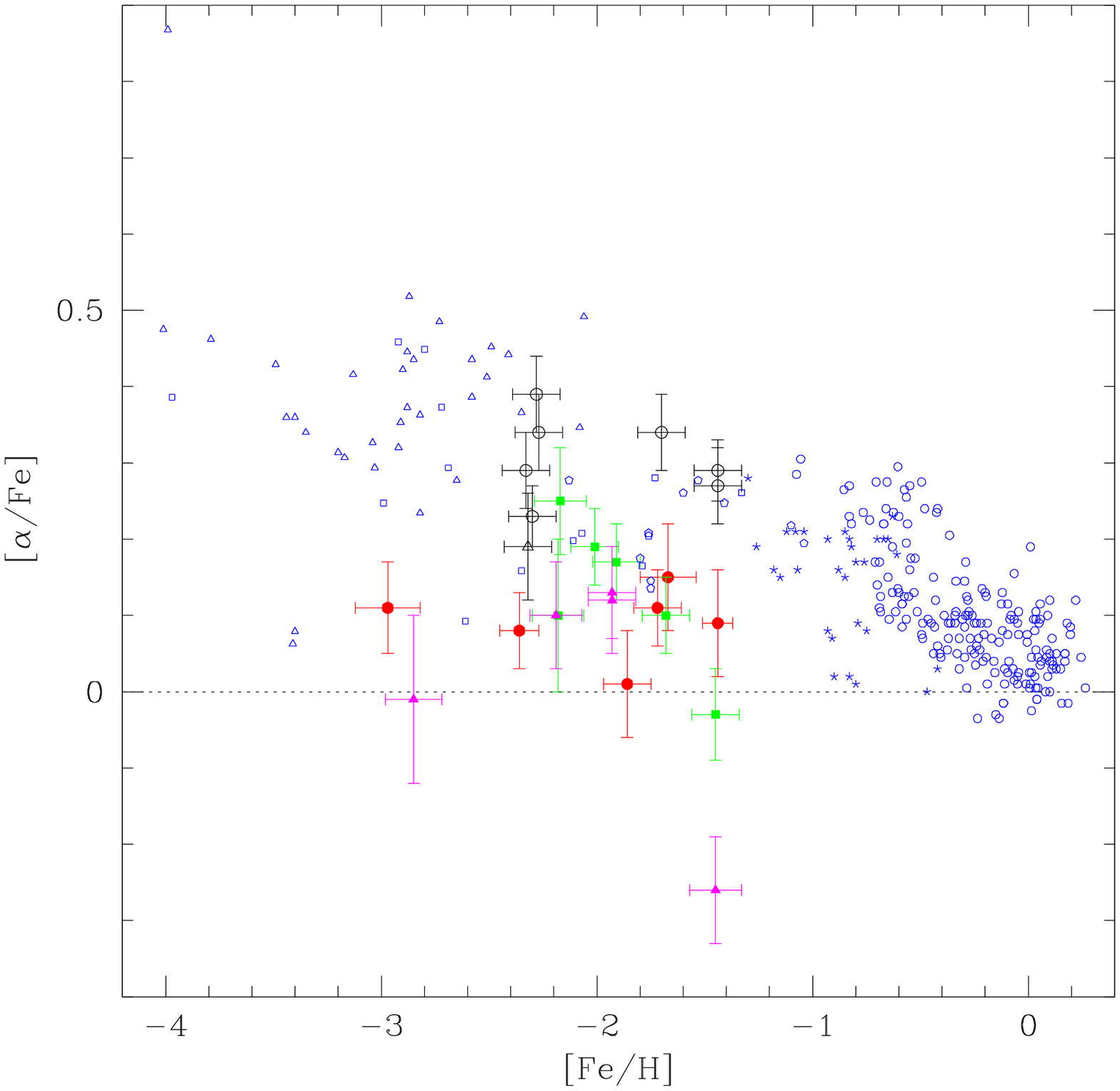} 
 \caption{Observed [$\alpha$/Fe] vs. [Fe/H] in the Galaxy and in dwarf spheroidals (data with error bars). The figure is from Shetrone et al. (2001).}
   \label{fig1}
\end{center}
\end{figure}

\section{Connection between Type Ib/c SNe and GRBs}
Some long GRBs have been found to be associated with Type Ib/c SNe, therefore it is interesting to check whether the Type Ib/c SN rates in galaxies are compatible with the observed GRB rate. Bissaldi et al. (2007) assumed that Type Ib/c SNe arise either from : i) single WR stars with masses $M> 25 M_{\odot}$ or from ii) massive stars (12-20 $M_{\odot}$) in binary systems. In both cases, in fact, a massive star explodes after having lost its H mantle thus resulting into a TypeIb/c SN. They also assumed different SFRs in different galaxies going from a short and intense burst in ellipticals to a low and continuous SF in irregulars. The agreement with the observed rate was found to be good for spirals and irregulars (ellipticals do not show Type Ib/c SNe). As a second step, they computed the cosmic Type Ib/c SN rate by assuming several cosmic SFRs. In particular, they tested  the cosmic SFR of Calura (2004) which assumes that ellipticals form very quickly and at high redshift as well as cosmic SFRs derived in the context of the hierarchical clustering scenario for galaxy formation, where ellipticals, especially the most massive ones formed last and until recently. Clearly the cosmic SFR of Calura predicts a high Type Ib/c rate at high redshift due to massive ellipticals, whereas the hierarchical cosmic rate is strongly decreasing at high redshift. 
In Fig. 6 we show the predicted cosmic Type Ib/c SN rates, under different assumptions about the cosmic SFR, compared with the observed GRB rate.
\begin{figure}[b]
\begin{center}
\includegraphics[width=3.4in]{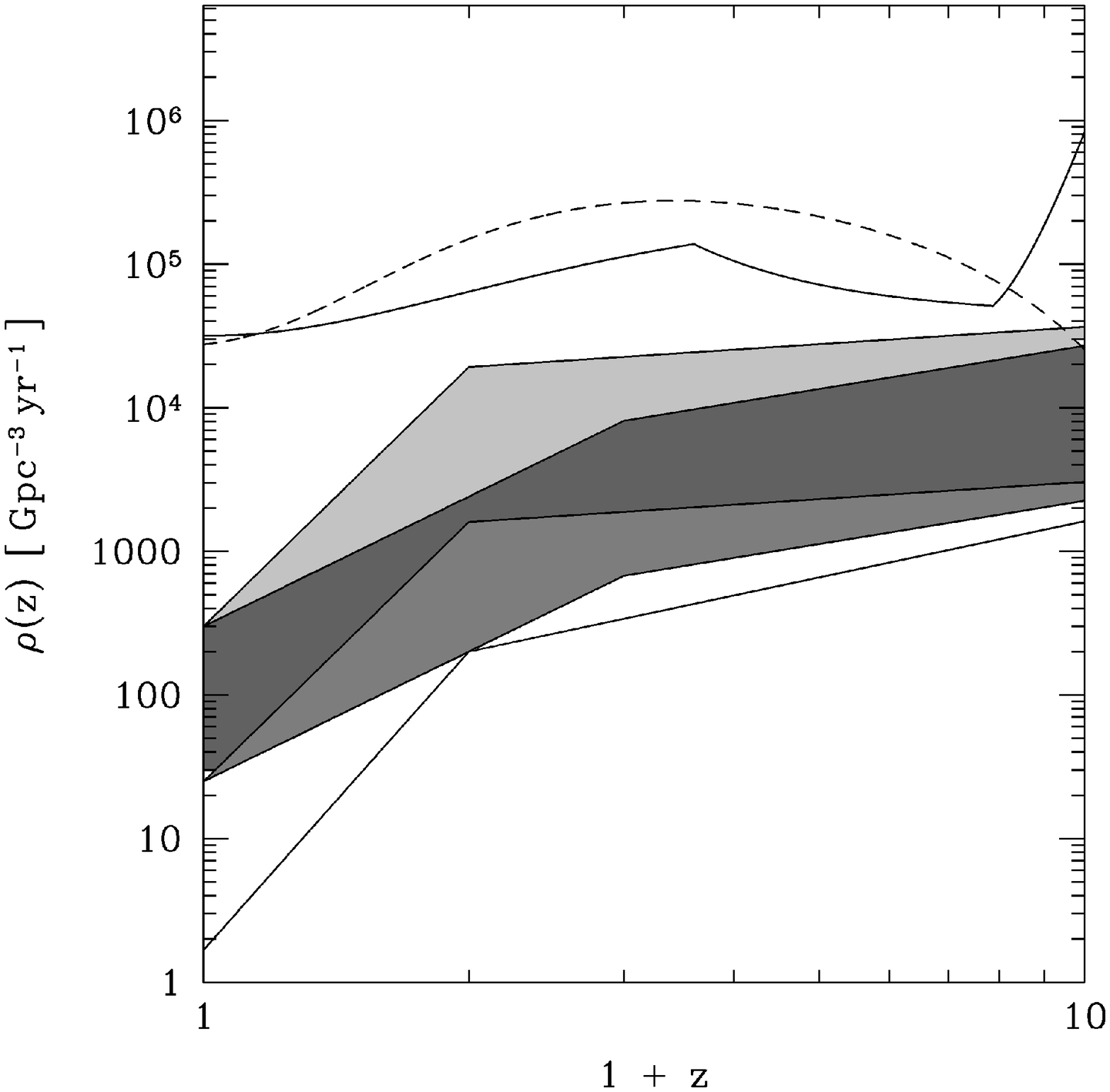} 
 \includegraphics[width=3.4in]{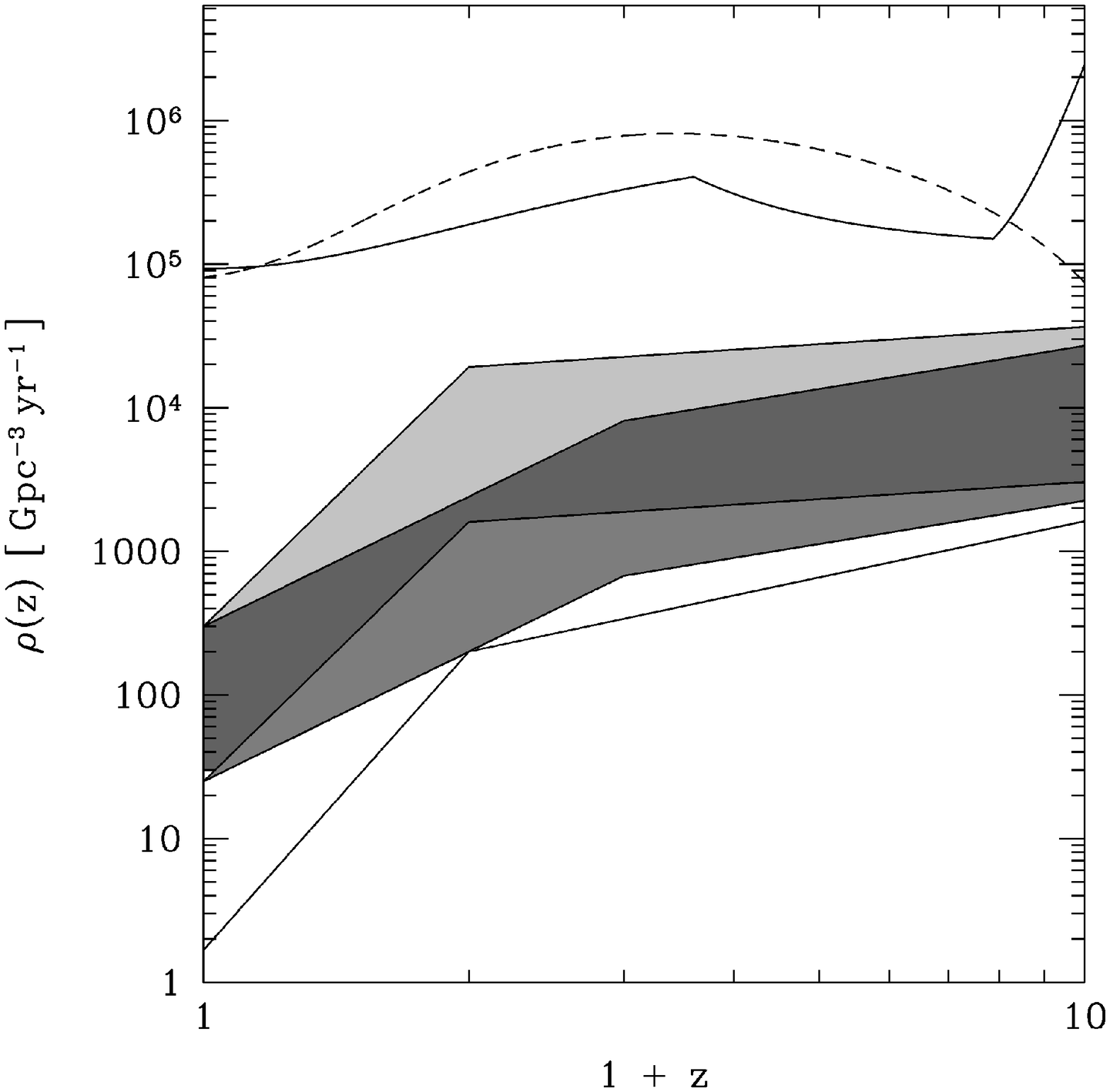} 

 \caption{Comparison between the observed cosmic GRB rate and the predicted cosmic Type Ib/c SN rates as functions of redshift.Upper continuous line: predictions from the cosmic SFR favoring high star formation in ellipticals at high redshift (Calura 2004). Dashed line: predictions from a hierarchical cosmic SFR. Shaded area and lowest continuous line represents the data relative to the GRB rate. Upper panel: predicted Type Ib/c SN rates assuming a Salpeter (1955) IMF. Bottom panel: predicted Type Ib/c SN rates assuming a top heavy IMF. Figure and references can be found in Bissaldi et al. (2007).}
   \label{fig1}
\end{center}
\end{figure}
\section{Conclusions}

We have discussed the importance of the time-delay model in interpreting the abundance patterns in galaxies. In particular, we have discussed the role of core-collapse SNe (Type II, Ib/c)  relative to the thermonuclear SNe (Type Ia) in the galactic chemical enrichment.
We can conclude the following:
\begin{itemize}
\item abundances in metal poor stars do not always show the signature of massive stars, it rather depends on the history of star formation. In fact, some low metallicity objects show the signature of Type Ia SNe (e.g. dwarf spheroidals) and this is interpreted as due to slow star formation. On the contrary, the stars in the Galactic bulge and spheroids are metal rich and do not show the signature of Type Ia SNe: this is interpreted as due to a very fast star formation process.

\item The [O/Mg] vs. [Mg/H] relation, both in the bulge and in the disk of the Milky Way, strongly favors a metallicity dependent mass loss in massive stars. The plot [N/O] vs. [O/H] in the Galaxy strongly favors the production of primary N from massive stars. This can be achieved by means of stellar rotation in massive stars.

\item Elliptical galaxies show an increasing average [Mg/Fe] ratio in stars with galactic mass: this strongly supports the idea that the most massive ellipticals formed stars for a shorter time than the less massive ones (downsizing).

\item By assuming that Type Ib/c SNe arise from either single massive stars ($M> 25 M_{\odot}$) or from massive stars in binary systems, we compared the cosmic Type Ib/c SN rate with the observed GRB rate. The Type Ib/c SN rate is much higher than the GRB rate which is only a fraction varying from 0.1 to 1\% of the Type Ib/c rate, in agreement with observational determinations (e.g. Della Valle 2005). We also predict that if we accept downsizing in star formation in ellipticals, then the number of GRBs at high redshift should be much higher than predicted in hierarchical models of galaxy formation.
\end{itemize}

\begin{discussion}

\discuss{Hirschi}{Can you differentiate between a lower minimum mass limit for single stars to form Wolf-Rayet stars and a larger fraction of binaries to reproduce the number of SNe Ib/c?}

\discuss{Matteucci}{Yes, we can in principle. There should be a lower limit for single WR stars which can reproduce the observed Type Ib/c SN rate but in this case one has also to check that other constraints are not violated. If we find a too small limiting mass for WR stars, incompatible with stellar evolution calculations, then we have a constraint. We will test this point in the future.}

\end{discussion}

\end{document}